# Anisotropic Etching and Nanoribbon Formation in Single-Layer Graphene

Leonardo C. Campos[1,3]*, Vitor R. Manfrinato[2,4]*, Javier D. Sanchez-Yamagishi[1], Jing Kong[2], Pablo Jarillo-Herrero[1]

[1]Department of Physics, Massachusetts Institute of Technology, Cambridge, MA 02139

[2]Department of Electrical Engineering and Computer Science, Massachusetts Institute of Technology, Cambridge, MA 02139

[3]Departamento de Física, Universidade Federal de Minas Gerais, Belo Horizonte, MG CEP: 31270-901, Brazil

[4]Departamento de Engenharia de Sistemas Eletrônicos, Escola Politécnica, Universidade de São Paulo, São Paulo, SP CEP: 05508-970, Brazil

**Abstract:** We demonstrate anisotropic etching of single-layer graphene by thermally-activated nickel nanoparticles. Using this technique, we obtain sub-10nm nanoribbons and other graphene nanostructures with edges aligned along a single crystallographic direction. We observe a new catalytic channeling behavior, whereby etched cuts do not intersect, resulting in continuously connected geometries. Raman spectroscopy and electronic measurements show that the quality of the graphene is resilient under the etching conditions, indicating that this method may serve as a powerful technique to produce graphene nanocircuits with well-defined crystallographic edges.

Graphene holds many exciting possibilities for demonstrating new physics as well as novel electronic applications[1, 2], many of which can only be realized by confining graphene into nanoribbons and other nanostructures. For example, ballistic room-temperature transistors[3-5] and carbon-based spintronic devices[6-10] are two tantalizing possibilities which could one day be realized in a graphene nanodevice. First though, a reliable method must be found to controllably produce graphene nanostructures with specific sizes, geometries, and defined crystallographic edges. Theoretical predictions indicate that a graphene nanoribbon with zig-zag edges can behave as a half-metal[6, 7] which, paired with graphene's long spin relaxation time[11], could be used to produced spin-valves and other spintronic devices. In addition, nanosized geometric structures such as triangles with zig-zag edges are predicted to have a net nonzero spin[8, 9, 12, 13], furthering the potential use of graphene as a canvas for spintronic circuits. For field effect transistor applications, quantum confinement induces a band gap in the normally gapless graphene[10, 14, 15], but the potential performance of the device depends strongly on the edge structure as well [4, 16, 17].

Graphene nanoribbons have already been produced using plasma etching[18-21], STM lithography[22], and AFM anodic oxidation[23], as well as by chemically derived techniques[24]. However, a method to produce single-layer graphene structures with well-defined crystallographic edges remains elusive. The hope for an anisotropic etching method is not unfounded though, because zigzag and armchair edges have markedly different chemical reactivities[25, 26]. Indeed, previous studies of catalytic gasification of carbon found that catalytic metal nanoparticles would sometimes etch graphite along crystallographic directions, creating both armchair and zigzag edges[27-31].

This effect has been studied more recently in the context of graphene applications [30-32], but there have been no results to date in isolated single-layer graphene.

Here we present a method to produce crystallographically-oriented cuts in single-layer graphene (SLG) which effectively has all cuts oriented with the same edge-chirality, in stark contrast to the same process in graphite. We also show that the cuts produced in SLG avoid crossing each other, resulting in continuously connected graphene nanostructures. Using this method we fabricate nanoribbons, equilateral triangles and other graphene nanostructures which could feature novel electronic behavior resulting from their specific edge orientations.

To prepare graphene samples for nanoparticle-assisted etching we follow multiple cleaning steps to ensure reliable production of cuts. Before depositing graphene, we clean $SiO_2$/Si substrates with acetone and isopropyl alcohol and submit them to UV illumination for 5 minutes to remove organic contamination. The UV illumination also has the important role of turning the $SiO_2$ surface more hydrophilic[33], which ensures proper wetting during the spin-coating deposition of Ni from solution. On these clean substrates the graphene is mechanically exfoliated using semiconductor grade tape. We identify SLG by a combination of: i) optical microscopy contrast[34], ii) atomic force microscopy (AFM) (SLG typically exhibits a 0.8nm height[35]), and iii) Raman spectroscopy (graphene's G' (2D) peak exhibits a single Lorentzian shape[36]). Next, to remove tape residue the graphene samples are heated in a quartz tube at 500°C for 15min under Ar:H2 flow (850:150 sccm). After the heat cleaning process, a solution of ($NiCl_2$:$H_2O$) at a concentration of 2.4mg/mL is spun at 1800RPM for 60s on to the substrate surface and then baked for 10 mins at 90° C on a hot plate to evaporate the $H_2O$. This $NiCl_2$ treated sample is then submitted to a 2 step process under Ar:$H_2$ flow (850:150 sccm): annealing at 500°C for 20 min resulting in Ni nanoparticle formation, and then etching at 1000°C for 25 min. Optimal etching conditions were found to require a slow heating/cooling rate of 50° C/min.

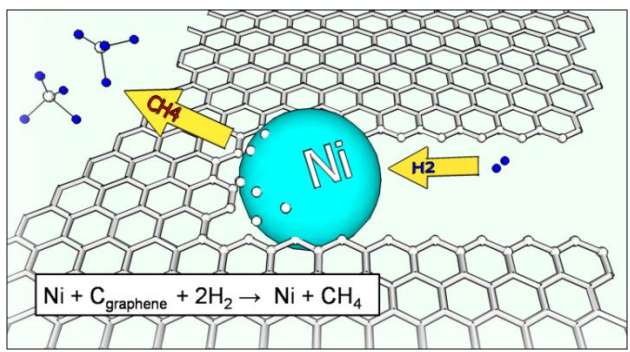

Figure 1. Cartoon of a Ni particle etching a graphene sheet (not to scale). Ni nanoparticles absorb carbon from graphene edges which then reacts with $H_2$ to create methane. (Inset) Summary of the hydrogenation reaction that drives the etching process.

During the high-temperature etching stage, the Ni nanoparticles etch the graphene through catalytic hydrogenation of carbon, where carbon atoms from exposed graphene edges dissociate into the Ni nanoparticle, and then react with $H_2$ at the Ni surface[29]. This process is summarized by the chemical reaction in Figure 1, where carbon from the graphene is hydrogenated into methane by the Ni nanoparticle catalyst. The same reaction can also be understood as the effective reverse of catalytic carbon nanotube (CNT) growth[37, 38], and indeed we found that the growth of CNTs competes with the desired etching results in graphene. This occurs because at high concentrations of carbon the Ni nanoparticles become super-saturated with carbon and can begin to expel carbon nanotubes[37, 38]. It is for this reason that we take extra precaution during preparation to avoid organic contaminants which can act as carbon sources and saturate the Ni nanoparticles. In addition, methane produced by the etching process itself can supersaturate the Ni nanoparticles if the amount of Ni on the substrate is too low. When the concentration of applied $NiCl_2$ solution is less than 0.05mg/mL, the typical result is formation of CNTs on the sample. On the other hand, when applying concentrations larger than 2.4mg/mL, which results in more Ni nanoparticle production, we observe no CNTs and only cuts on the graphene and graphite (see supplementary materials). The higher concentration of Ni also results in more cuts produced on the graphene surface, but we find that this density increase can be balanced by reducing the annealing time if a lower density of cuts is desired.

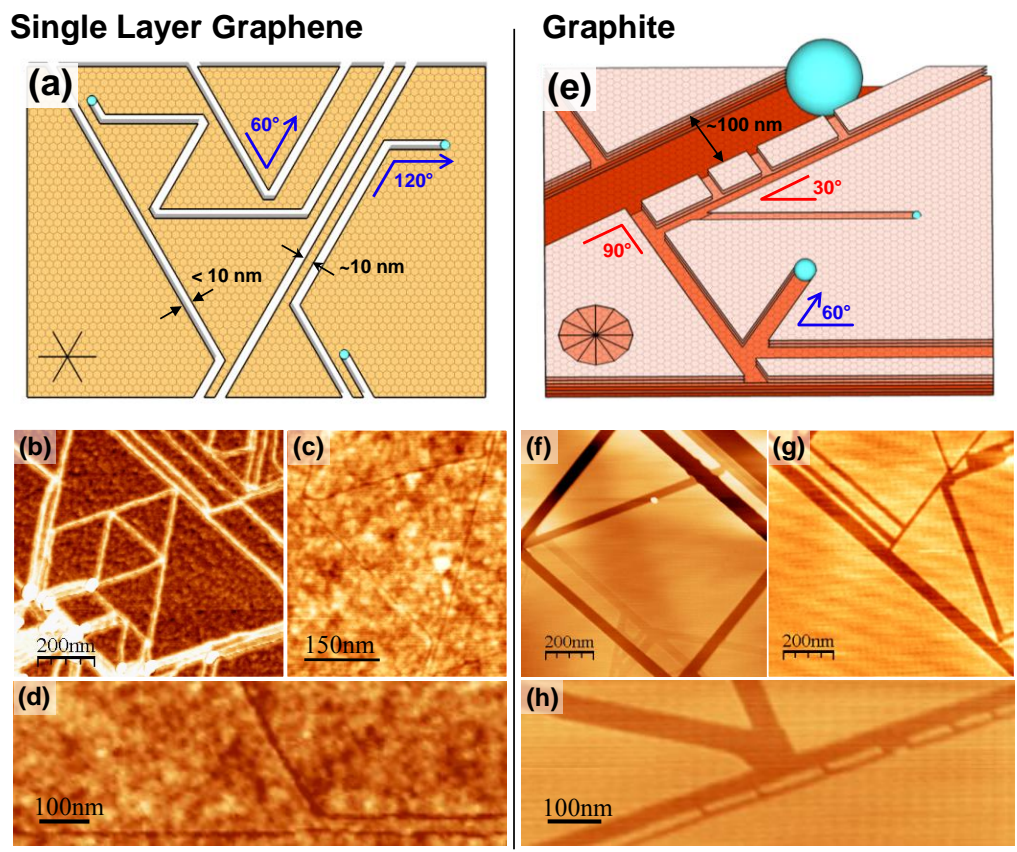

Figure 2. Comparison of nanoparticle-assisted etching in SLG and graphite **(a) Key features of etching in SLG are**: chirality-preserving angles of 60° and 120°; avoided crossing of trenches leaving ∼10nm spacing between adjacent trenches and producing connected nanostructures; trenches and nanoparticles with uniform width < 10nm. (b) AFM Phase image of etched SLG with produced geometric nanostructures. The phase image obscures small details, making adjacent trenches appear to merge together. (c) AFM height image of equilateral triangle connected to 3 nanoribbons. (d) AFM height image of a trench which avoids crossing another trench, running parallel to it. (c,d color scale 0 to 1.7 nm) **(e) Key features of etching in Graphite and Few-Layer Graphene are:** Chirality-changing angles of 90°, 150° and 30°, in addition to 60° and 120°; trenches which merge, producing disconnected geometries; trenches and nanoparticles of varying size (10-1000nm). (f-h) AFM height images of etched graphite showing the previously mentioned features. (color scale 0 to 7 nm)

The etching process produces a mosaic of clearly defined cuts across the SLG surface, as measured by AFM (Fig 2b-d). These continuous trenches left behind by individual nanoparticles run along straight lines, intermittently deflecting from their path or reflecting away from previously etched trenches. These deflections and reflections show a surprising regularity, with measured angles between any pair of trenches of either 60° or 120° (Fig 2a). As can be seen in Figure 3, trenches forming edges at angles of 60° and 120° preserve the chirality of the edges, indicating that nearly all the cuts in our samples run along the same crystallographic orientation. A further striking feature is that when an etching nanoparticle approaches within ~10nm of a previously etched trench, it turns away before reaching the trench (Fig 3). This surprising fact has been verified by all high resolution

AFM images, and is what allows for the creation of connected graphene nanostructures such as crystallographically-oriented nanoribbons. For example, Figure 3a shows a trench that approaches another trench, only to turn and run parallel to it, creating a long sub-10nm wide nanoribbon. Eventually, the trench turns away again, forming 2 large micron-size areas of graphene connected to the nanoribbon to which electronic contacts can be attached. Using this method we have been able to produce nanoribbons from a few nanometers to micrometers in length, as well as other exotic structures such as triangles and parallelograms.

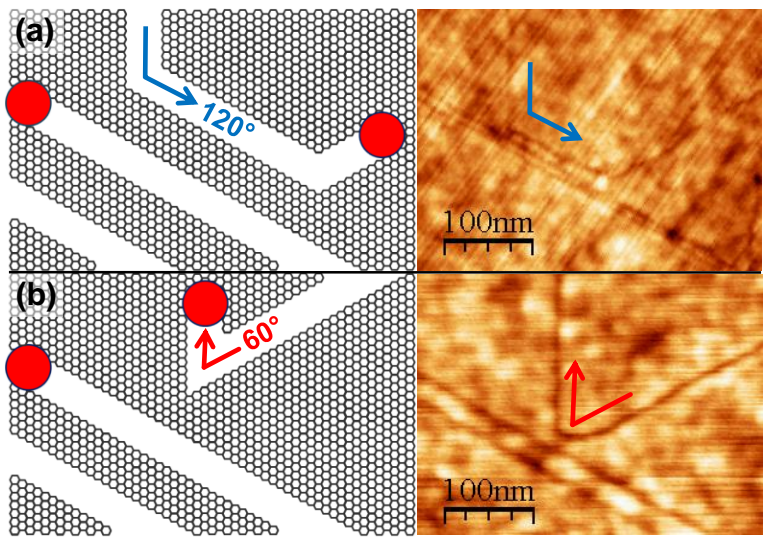

Figure 3. Creation of crystallographically-oriented nanoribbons. Nanoparticles approach previously etched trenches and turn, forming angles of 60° or 120°. (a) Drawing and AFM image of a nanoribbon created by a nanoparticle which turns twice, forming two angles of 120°. (b) Drawing and AFM image of a nanoconstriction formed by a nanoparticle which turns sharply near another trench to form a 60° angle.

We emphasize that these key constraints on nanoparticle etching are unique to SLG, in contrast to the less constrained behavior found in our own measurements and previously reported results in graphite and few-layer-graphene [30, 31]. As seen in Figure 2, trenches cut in graphite usually meld when they approach each other or even cross, where as in graphene the trenches are kept separate. Hence, the process that produces connected nanoribbons and other nanostructures in SLG does not occur in graphite samples. In addition, while cuts in graphene generally form angles of 60° and 120°, in graphite it is possible to see angles of 30°, 90° and 150° which indicate changes of the crystallographic orientation of the trenches, and hence a change in the chirality of the edges produced. To study this behavior more in depth, we measured over 200 trenches across 10 different samples and counted the angles of deflections and reflections of single continuous trenches (i.e. clearly etched from the same nanoparticle). Through this measurement we were able to verify that indeed, chirality changing angles of 90° and 150° are common in graphite, but exceedingly rare in graphene. In SLG we measured 98% of the angles formed by the cuts to be either 60° or 120° (Fig 4b), in other words, the general chirality (zigzag or armchair) of the trench edges is preserved for all the cuts in etched SLG. On the rare occasion where we observed an angle of 90° or 150° in SLG, we found that these turns were unstable and would quickly turn again, reverting to the original crystallographic orientation. This is further evidence that etching along a specific crystallographic orientation is favored in SLG.

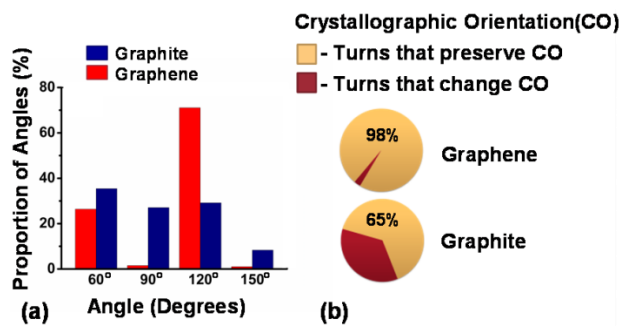

Figure 4. Statistics for angles formed by turns in individual trenches in graphene and graphite. (a) Distribution of angles formed by turns. (b) Proportion of turns that preserve the crystallographic orientation of a trench. Trenches that form angles of 60° and 120° will have the same crystallographic orientation, trenches at angles of 90° and 150° will have different crystallographic orientations and correspondingly different edge-chiralities.

The direct measurement of the crystalline structure of the trench edges has not been performed yet, but previous TEM studies of Ni nanoparticle etching in graphite found that Ni nanoparticles tended to etch along graphite's <11-20> crystallographic direction[28, 29], which corresponds to the production of zig-zag edges. Recent ab initio calculations also show that the removal of carbon atoms from armchair edges, thus producing zig-zag edges, is energetically more favorable than removing carbon atoms from zigzag edges[31]. The question of edge chirality could also be related to why nanoparticles etching through SLG do not etch into previously produced trenches  For example, Coulomb interactions between the nanoparticles (charged due to the graphene-Ni work function difference) and the enhanced electronic density of states at the zigzag edge[15] may prevent the nanoparticle from intersecting the previously etched trench. However, this phenomenon requires quantitative theoretical modeling beyond the scope of this paper.

A final characteristic in graphene that lends to more regular etching patterns is that all trenches in SLG have widths less than 12nm, while in graphite, trenches from 10nm to micrometers in width are observed. This may be because graphite's multi-layer edge supports the formation of larger nanoparticles, while only small nanoparticles form on the SLG edge during the annealing process. With AFM it is difficult to measure the precise width of the narrow trenches found in SLG, but the average height of the nanoparticles that produce these trenches is accurately measured at 4 nm, which is consistent with the size of the trenches produced.

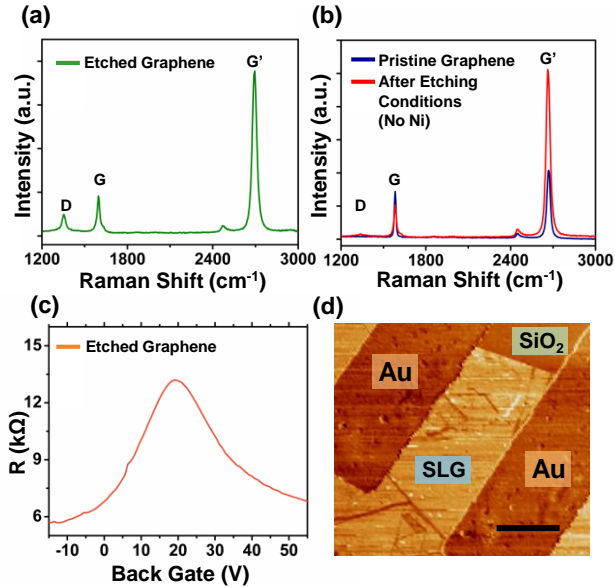

Figure 5. Raman spectrum measured at wavelength 532nm and electronic measurements. (a) Raman shift for a single-layer graphene flake after etching with Ni nanoparticles. The D peak at 1350cm$^{-1}$ is proportional to defects in the graphene crystal structure. (b) Raman shift of pristine SLG (blue line) and the same SLG after being submitted to etching conditions without Ni nanoparticles (red line). (c) Resistance vs. voltage measurement of an etched graphene sample with a Dirac resistance peak typical for pristine SLG. (d) AFM phase image of etched graphene sample with electrodes (1μm scale bar).

Given the high temperatures used during this etching process, the preservation of the quality of the graphene becomes an important concern. To determine the quality of the graphene after the etching process we performed a combination of Raman spectroscopy and room temperature electrical measurements. Figure 5b (blue line) shows the typical Raman spectrum of a pristine SLG sheet, with labeled peaks[36, 39, 40]: G at 1580cm$^{-1}$ and G' (also called 2D) at 2700cm$^{-1}$. The main difference found in the Raman spectrum of graphene before and after etching is the presence of the D peak centered at 1350cm$^{-1}$ in etched graphene (Fig 5a). The D peak in the Raman spectra originates from defects in the graphene crystal structure such as missing carbon atoms and adsorbed atoms, and from armchair edges in the structure[40-44]. From this spectrum alone though, it is not clear whether the D peak is caused by the presence of edges from the etching process or from other defects caused by the high temperature process. For further investigation we submitted a sample to all the conditions of the etching process, but without Ni nanoparticles. As is shown in Figure 5b, the graphene has no D peak before the etching conditions (blue line), and almost no change after (red line). This indicates that heating of the sample under etching conditions alone is not responsible for the D peak after etching, since the D peak after the full etching process is more than an order of magnitude larger than the D peak for the sample only submitted to the etching conditions. Theoretically, only edges of armchair type should produce a D peak[41]; but it is not possible to identify the chirality of the cut edges from this measurement since a quantitative description of the relationship between edge disorder and D peak intensity has not been established.

Electronic measurements of the etched SLG were also performed to compare with the electronic behavior of pristine graphene flakes. Two contacts were placed in a region with a low density of cuts to avoid electrical separation of the contacts (Fig 5d), and the resistance as function of gate voltage was measured. Figure 5c shows the resulting typical Dirac resistance peak, similar to those observed for pristine graphene samples[1]. From these measurements we conclude that the etching process preserves the general electrical characteristics of graphene, verifying that this method is viable for producing graphene nanostructures useful for future experimentation.

In summary, we have demonstrated anisotropic etching in single-layer graphene which produces connected graphene nanostructures with crystallographically-oriented edges. This opens many future avenues to study graphene nanostructures such as nanoribbons, nanoconstrictions and quantum dots with crystallographic edges.

**Supporting Information Available:** Further details on the etching process and the growth of carbon nanotubes can be found in the Supporting Materials. This material is available free of charge via the Internet at http://pubs.acs.org

## Acknowledgements

We thank H. Farhat for help with Raman spectroscopy. L.C.C. acknowledges support by the Brazilian agency CNPq. V.R.M acknowledges the Roberto Rocca Education Program. J.D.S-Y acknowledges support by NSF GRFP. This work made use of the MRSEC Shared Experimental Facilities supported by NSF (DMR – 0819762).